\documentclass[superscriptaddress,floatfix,showpacs,aps,twocolumn]{revtex4-2}
\usepackage{graphicx,amsfonts,amsmath,bm,xcolor,tabularx,epstopdf}
\usepackage[utf8]{inputenc}
\usepackage[english]{babel}
\usepackage{amssymb}
\usepackage{hyperref}
\usepackage{wrapfig}

\usepackage[rightcaption]{sidecap}

\usepackage[normalem]{ulem}

\newcommand{\I}{{\rm i}}
\begin{document}

\title{Quantum magnetic phase transitions in a Kugel-Khomskii model including spin–orbit coupling}

\author{D. E. Chizhov, P. A. Igoshev, V. Yu. Irkhin}

\affiliation{M. N. Mikheev Institute of Metal Physics, Ekaterinburg, 620108 Russia}
%\date{\today}
\begin{abstract}
Using the formalism of pseudospin and isospin operators the Hamiltonian of an effective Kugel-Khomskii model with spin–orbit coupling is derived 
 with an exact account of the t$_{2g}$ multiplet splitting by the crystal field. An analytical solution  is obtained for an  arbitrary  relation between the Hubbard repulsion and crystal field splitting, i.e., interpolating the cases of Mott-Hubbard and charge-transfer insulators. A description of orbital orders is given in terms of octupole moments.   The ground-state phase diagram is constructed in the parameter space spanned by spin-orbit coupling, Hund's exchange, and Hubbard interaction. We investigate a quantum phase transition between states exhibiting hidden magnetic and orbital long-range order and a ferromagnetic state with a reduced magnetic moment accompanied by antiferroorbital order. It is shown that the cooperative effect of Hund’s   and spin–orbit interactions  gives rise to an easy-plane-type anisotropy. 
%A finite value of the crystal field parameter induces an additional  easy-axis anisotropy and anisotropy in the $xy$ plane. %within the plane of the square lattice.??
\end{abstract}

\maketitle

\section{Introduction}

Anisotropic magnetic systems demonstrate a variety of  phase diagrams including non-trivial magnetic transitions in the ground state. In particular, such transitions occur in the case of easy-plane anisotropy \cite{egami1975theory,val1982hubbard,val1983contribution,onufrieva1988field,irkhin1998quantum}.
%T. Egami and S.S. Brooks, Phys. Rev. B12, 1029 (1975)  F.P. Onufrieva, Fiz.Tverd.Tela 23, 2664 (1981); Zh.Eksp.Teor.Fiz. 95, 899 (1989)   V.V. Valkov and S.G. Ovchinnikov, Teor. Mat. Fiz., 50, 466 (1982); Zh.Eksp.Teor.Fiz. 85, 1666 (1983)  V. Yu. Irkhin and A. A. Katanin, Phys. Rev. B 58, 5509 (1998).
 Usually such a consideration is performed within simple phenomenological Heisenberg model. At the same time, occurrence of magnetic anisotropy is intimately related to the spin–orbit coupling.
A number of  strongly correlated systems with orbital degeneracy investigated recently possess orbital degrees of freedom and strong spin–orbit coupling, which leads to unquenching of the orbital momenta.
%Явление дальнего орбитального порядка, формирующегося в сильнокоррелированных системах с орбитальным вырождением, привлекает внимание исследователей, начиная с 1970-х годов~\cite{kugel1982jahn}. 
Magnetic order in  such systems is determined by the competition of various types of interactions: spin-orbit coupling, crystal field splitting, and Hund exchange. The latter is the cause of the orbitally alternating ferromagnetic state with the maximum order parameter. 

Magnetic phase transitions in Mott insulators with $d^4$ electron configurations were studied with account of  Hund’s interaction  and spin–orbit coupling in Refs. \cite{meetei2015novel,feng2020magnetic}.
Another example of a relevant system is the antiferromagnetic $d^1$ insulator Sr$_2$VO$_4$ with a perovskite structure. 
%containing one valence $d$-electron per site (a $d^1$-configuration of the $t_{\rm 2g}$-multiplet).
This compound exhibits properties of both an antiferromagnet and a weak ferromagnet.
%при  температурах ниже $100$~K~\cite{nozaki1991layered}. 
%Для объяснения оптических~\cite{matsuno2005variation} и магнитных~\cite{zhou2007orbital} свойств была предложена идея о существовании орбитального порядка в плоскости $\rm VO_2$ при температуре $\lesssim 120$~K. 
Magnetic susceptibility and heat capacity measurements demonstrate an orbital order below 100 K and  transition to a  ferromagnetic state with reduced moment at $T < 10$ K \cite{Viennois2010, Sakurai2015}.
%Neutron scattering experiments indicate the existence of hidden magnetic order formed by Kramers doublets at low temperatures \cite{zhou2010orbital}.
%Эксперименты по мюонной спиновой спектроскопии ($\mu$SR) свидетельствуют о формировании немагнитного дальнего порядка при $T \lesssim 100$~K и  антиферромагнитного порядка ниже 8~К~\cite{Sugiyama2014} или неоднородного ферромагнитного состояния~\cite{yamauchi2015structural}. 

%A number of pictures have been proposed to explain the observed properties of Sr$_2$VO$_4$: a dimerized solution of an effective Kugel-Khomskii model with direction-dependent exchange integrals~\cite{jackeli2007dimer}, a two-sublattice long-range order corresponding to the maximum of the projection of the total electron moment~\cite{Eremin2011}, a two-sublattice order of octupoles or isospins stabilized by strong spin-orbit coupling~\cite{Jackeli2009a}.

In the present paper, we treat a microscopic model with orbital degeneracy  and anisotropy, which generalizes the famous Kugel-Khomskii model~\cite{kugel1982jahn}. We provide an analytical solution in the two-sublattice approximation  depending on the parameters of the  spin-orbit coupling and  Hund's interaction.
%in the case of a large splitting by the crystal field between the $xy$- and $xz, yz$-states. 

%%%%%%%%%%%%%%
% Искуственный сдвиг текста вверх:
%\vspace{-2.0cm}
%%%%%%%%%%%%%%
\section{The model}

Our starting point is the degenerate Hubbard model for the $t_{2g}$ orbitals on a square lattice:
\begin{equation}\label{eq:H_full}
\mathcal{{H}} = \mathcal{{H}}_{\rm local}  + \mathcal{{H}}_{\rm tr} ,
\end{equation}
where
\begin{equation}\label{eq:H_local}
   \mathcal{{H}}_{\rm local} = \mathcal{{H}}_{\rm Coulomb} + \mathcal{{H}}_{\rm CF} + \mathcal{{H}}_{\rm so}. 
\end{equation}
Here
$\mathcal{{H}}_{\rm tr} = \sum_{<i,j>,m,\sigma}t^{m}_{ij}c^\dag_{im\sigma}c^{}_{jm\sigma}$
describes the transfer of electrons between the sites $i, j$, $m= xz,xy,yz$  and $\sigma = \uparrow,\downarrow$ are spin and orbital indices of the electron creation and annihilation operators,
$$t^m_{ij} = -t\left(a^{xz}_{ij}\left(\delta_{m,xy}+\delta_{m,xz}\right)+a^{yz}_{ij}\left(\delta_{m,xy}+\delta_{m,yz}\right)\right)$$ are hopping integrals, $a^{xz}_{ij}$ $\left(a^{yz}_{ij}\right) =1$, provided that $i$ and $j$ are the nearest neighbors along the $x$-axis $\left(y\right)$, otherwise it is zero.
$\mathcal{{H}}_{\rm CF} = \Delta_{\rm{CF}}\sum_{i\sigma}c^\dag_{i,xy,\sigma}c^{}_{i,xy,\sigma}$ describes the crystal field splitting between the one-electron level $xy$ and the doublet $xz,yz$, $\mathcal{{H}}_{\rm so} = -\lambda\sum_{i,m,m'}\mathbf{s}_{i}\cdot\mathbf{l}_{imm'}$ is spin-orbit coupling, $\lambda\ge0$,
$\mathbf{s}_i=\frac12\sum_{m\sigma\sigma'}c^\dag_{im\sigma}\boldsymbol{\sigma}_{\sigma\sigma'}c^{}_{im\sigma'}$,
where
$\boldsymbol{\sigma}$ is the vector of the Pauli matrices.
The effective orbital-moment  Cartesian $o$-components $l_{imm'}^{o} = -\I\sum_{\sigma} l^{o}_{mm'}c^\dag_{im\sigma}c^{}_{im'\sigma}$ are constructed from matrix elements of the effective angular momentum $l = 1$~\cite{streltsov2017orbital}: $l^{o}_{mm'} = -\I \varepsilon_{\tilde omm'}$ where $\varepsilon$ is the Levi-Civita tensor,
$\tilde x = yz, \tilde y = xz, \tilde z = xy$.

%%%%%%%%%%%%%%
% Искуственный сдвиг текста вверх:
%\vspace{-2.0cm}
%%%%%%%%%%%%%%
%$\mathcal{{H}}_{\rm Coulomb}$ описывает локальное кулоновское взаимодействие электронов. 
For $t_{\rm 2g}$ states, the on-site Coulomb (Hubbard) interaction operator $\mathcal{{H}}_{\rm C}$ reads:
\begin{multline}
\mathcal{{H}}_{\rm C} = \frac{U}{2} \sum\limits_{i,m,\sigma}n_{im\sigma}n_{im-\sigma} + \frac{U'}{2} \sum\limits_{i,m\neq m'}n_{im}n_{im'} \\ 
- \frac{J}{2} \sum\limits_{i,m\ne m';\sigma,\sigma'}c^\dag_{im\sigma}c^{}_{im\sigma'}c^\dag_{im'\sigma'}c^{}_{im'\sigma} \\
-\frac{J_d}{2} \sum\limits_{i,m\ne m';\sigma}c^\dag_{im\sigma}c^{}_{im'-\sigma}c^\dag_{im-\sigma}c^{}_{im'\sigma},  
\end{multline}
where $U$ and $U'$ are Hubbard constants, $J$ is Hund's exchange  parameter, $J_d$ is pairwise many-electron hopping parameter. 
%$\bar\uparrow = \downarrow, \bar\downarrow = \uparrow$.

In the case where there is one electron per site, second order perturbation theory in hopping (with $\mathcal{H}_{\rm local}$ being a zero Hamiltonian)    yields the following effective Heisenberg  Hamiltonian
~\cite{2023:Igoshev_Streltsov_Kugel_JMMM,2024:Igoshev_Chizhov}:
\begin{equation}\label{eq:H_kin_presentation}
\mathcal{H}_{\rm eff} = \mathcal{H}^{\rm d}_{\rm eff} +  \mathcal{H}^{\rm p}_{\rm eff}, 
\end{equation}
where the term originating from virtual hopping to the wave function subspace of doubles ($\rm d$) reads
\begin{multline}\label{eq:H^d_start}
\mathcal{H}^{\rm d}_{\rm eff} = -\sum_{t=0,\pm} \frac{1/2-2\mathbf{s}_i\cdot\mathbf{s}_j}{E^\mathrm{d}_t}
\sum_{ijm_1m_2}t^{m_1}_{ij}t^{m_2}_{ji}\alpha^t_{m_1}\alpha^t_{m_2}
\\
\times
X_i^{m_1m_2}X_j^{m_1m_2}, 
\end{multline}
where 
\begin{equation}\label{eq:X_def}
X_i^{mm'} = \sum_\sigma c^\dag_{im\sigma}c^{}_{im'\sigma}
\end{equation}
is a Hubbard-like orbital transfer operator
and the eigenvalues and wavefunctions of the local Hamiltonian \eqref{eq:H_local} are restricted to  the invariant subspace of doubles $E^{\rm d}_0 = U - J_{\rm d}$, $\alpha^0_{xz} = -\alpha^0_{yz} = \frac1{\sqrt2}$,  $\alpha^0_{xy}  = 0$,
$$E^{\mathrm{d}}_s = U + \Delta_{\rm CF} + J_{\rm d}/2 + s \sqrt{9J_d^2/4 - J_d\Delta_{\rm CF} + \Delta_{\rm CF}^2},$$ 
$\alpha^s_{xz} = \alpha^s_{yz} = \mu$, $s = \pm$, $\mu_s$ being a normalization factor,
$$\alpha^s_{xy} = \mu_s\frac{E_s^{\rm d} + J_{\rm d} - U}{E_s^{\rm d} + J_{\rm d} - U - 2\Delta_{\rm CF}}.$$ 
 The term originating from pair-orbital hopping reads
\begin{multline}\label{eq:H^p.final.diagonal}
\mathcal{H}^{\mathrm{p}}_{\rm eff} = -\sum_{ij,n\ne n'}\frac{t^{n}_{ij}}{E^{\rm p}_{nn';\mathrm{T}}}
\left(
\left(1 - \frac{a_{nn'}}2\right)t^{n}_{ji}
X_i^{nn}X_j^{n'n'} 
\right.
\\
\left.
-  \frac{1+a_{nn'}}2t^{n'}_{ji}
X_i^{nn'}X_j^{n'n}  
-
2\mathbf{s}_i\cdot\mathbf{s}_j\left((1 - a_{nn'})t^{n'}_{ji}
X_i^{nn'}X_j^{n'n} 
\right.\right.
\\
\left.\left.
-
a_{nn'} t^{n}_{ji}
X_i^{nn}X_j^{n'n'}\vphantom{\int}\right)
\right),
\end{multline}
where $a_{nn'} = J/E^p_{nn';\mathrm{S}}$. 
Here $E^p_{nn';\mathrm{S,T}} = U' \pm J + \Delta_{\rm CF}(\delta_{n,xy} + \delta_{n',xy})$ are triplet/singlet eigenvalues of local Hamiltonian constructed from electron states on different orbitals $n$ and $n'$. 
%For brevity we introduce pair Hubbard-like exchange operators
%\begin{equation}
%\Pi^\pm_{ij}(m_1m_2;m_1'm_2') = \mathcal{S}^{(0)}_{i,m_1m_2}\mathcal{S}^{(0)}_{j,m_1'm_2'} \pm \bm{\mathcal{S}}^{}_{i,m_1m_2}\bm{\mathcal{S}}^{}_{j,m_1'm_2'}.
%\end{equation}

We treat the case of large $\Delta_{\rm CF}$. There are two regimes here (i)~$t^2/(U - 3J)\ll\Delta_{\rm CF} < U - 3J$ --- the regime of charge-transfer insulator where both Hubbard's $U$ and $\Delta_{\rm CF}$ play a role; (ii) $U - 3J \ll \Delta_{\rm CF}$ -- the Mott-Hubbard regime. 

In both these cases $\Delta_{\rm CF}$ is much larger than the relevant interspin exchange energy of order of $t^2/U$. Thus we exclude $xy$ state projecting the above Hamiltonian on the subspace spanned onto $xz$ and $yz$ states, which is marked by the bar symbol. 
To obtain compact results we consider  the regimes $\Delta_{\rm CF} \ll  U$ (so that $\Delta_{\rm CF}$ can be neglected in denominators) and  $\Delta_{\rm CF} \rightarrow +\infty$ (the strong Mott-Hubbard insulator regime). 
\begin{equation}
    \bar{\mathcal{H}}_{\rm eff} = \bar{\mathcal{H}}^{\mathrm{d}}_{\rm eff} + \bar{\mathcal{H}}^{\mathrm{p}}_{\rm eff}.
\end{equation}

In  both these regimes large $\Delta_{\rm CF}$ values result in vanishing of terms with $n = xy$ or $n' = xy$ in Eq.~\eqref{eq:H^p.final.diagonal}, so that $n = xz, n' = yz$ and vice versa.  The resulting Hamiltonian
\begin{multline}\label{eq:H^p_proj_xzyz}
\bar{\mathcal{H}}^{\mathrm{p}}_{\rm eff} = 
-\frac1{U'-J}\sum_{ij}\sum_{n\ne xy}t^{n}_{ij}
\\
\left(\vphantom{\int}\left(1 - \frac{a_0}2\right)t^{n}_{ji}
X_i^{nn}X_j^{\bar{n}\bar{n}} - \frac{1+a_0}2t^{\bar{n}}_{ji}
X_i^{n\bar{n}}X_j^{\bar{n}n}
\right.
\\
\left.
-
2\mathbf{s}_i\cdot\mathbf{s}_j\left((1 - a_0)t^{\bar{n}}_{ji}
X_i^{n\bar{n}}X_j^{\bar{n}n} - a_0t^{n}_{ji}
X_i^{nn}X_j^{\bar{n}\bar{n}} 
\vphantom{\int}\right)\right)
\end{multline}
does not depend on $\Delta_{\rm CF}$ in both regimes,  $\bar{xz} = yz$ and $\bar{yz} = xz$, $a_0 = a_{xz, yz}$. 
After projection we have $\mathcal{H}^{\rm d}_{\rm eff}\rightarrow\bar{\mathcal{H}}^{\rm d}_{\rm eff}$ with
%\begin{multline}
%\bar{\mathcal{H}}^{\mathrm{p}}_{\rm eff} =
%-\frac2{U'-J}\sum_{ij}\sum_{n\ne xy}
%\left(t^{n}_{ij}t^{n}_{ji}\Pi^+_{ij}(n,n;\bar{n},\bar{n})
%- 2t^{\bar{n}}_{ij}t^{n}_{ji}\mathcal{S}^{(0)}_{i;\bar{n}n}\mathcal{S}^{(0)}_{j;n\bar{n}}\right)
%\\
%+
%\frac{2J}{{U'}^2 - J^2}
%\sum_{n\ne xy}\sum_{ij}
%\left(t^{n}_{ij}t^{n}_{ji}\Pi^-_{ij}(n,n;\bar{n},\bar{n})
%+
%t^{n}_{ij}t^{\bar{n}}_{ji}
%\Pi^-_{ij}(n,\bar{n};\bar{n},n)
%\right).
%\end{multline}
%Wave functions of $\mathcal{H}^{\mathrm{d}}_{\rm eff}$ are the same, $\alpha^0_{xz(yz)} = -1/\sqrt2$, $\alpha^-_{xz,yz} = +1/\sqrt2$. However, eigenvalues of $\mathcal{H}^{\mathrm{d}}_{\rm eff}$ behave differently in these limits: $E^-_{\rm p}(\Delta \rightarrow 0) = U' - J_{\rm d}$, $E^-_{\rm p}(\Delta \rightarrow +\infty) = U' + J_{\rm d}$.  
%To take both these limits in the similar way we write
%where $\kappa = +1$ for the case i) and $\kappa = -1$ for the case ii) and 
\begin{multline}
\bar{\mathcal{H}}^{\rm d}_{\rm eff} = 
 -\frac2{U - J_{\rm d}}\sum_{ij}(1/4-\mathbf{s}_i\cdot\mathbf{s}_j) 
 \\
 \times \sum_{m_1m_2\ne xy}t^{m_1}_{ij}t^{m_2}_{ji}X_i^{m_1m_2}X_j^{m_1m_2}
 \left(\delta_{m_1m_2} - \frac{J_{\rm d}}{U + \beta J_{\rm d}} \right).
\end{multline}
Here $\beta = 2$ in the case  $J_{\rm ex}\ll\Delta_{\rm CF} \ll U - 3J$ (the charge-transfer regime, which is more realistic for systems like Sr$_2$VO$_4$) and $\beta = 1$ in the case $\Delta_{\rm CF} \gg U - 3J$ (the Mott-Hubbard regime).

For negative $\Delta_{\rm CF}$, $J_{\rm ex}\ll|\Delta_{\rm CF}|\ll U - 3J$ we perform a shift of the one-electron energy by $|\Delta_{\rm CF}|$ and get after projection onto the state $xy$ 
\begin{equation}\label{eq:H^d_tilde}
\tilde{\mathcal{H}}^{\rm d}_{\rm eff} = -\frac2{U^*}
\sum_{ij}|t^{xy,xy}_{ij}|^2(1/4 - \mathbf{s}_i^{xy}\cdot\mathbf{s}_j^{xy}),  
\end{equation}
with effective $1/U^* = \frac23(U - J_{\rm d})^{-1} + \frac13 (U + 2J_{\rm d})^{-1}$.
%and $\tilde{\mathcal{H}}^{\rm p}_{\rm eff} = 0$. 

In the case $\Delta_{\rm CF} < 0$, $|\Delta_{\rm CF}|\gg U - 3J $ we have the result \eqref{eq:H^d_tilde} with 
the replacement $U^*\rightarrow U$,
%\begin{equation} \tilde{\mathcal{H}}^{\rm d}_{\rm eff} = -\frac{2}{U} \sum_{ij}|t^{xy,xy}_{ij}|^2(1/4 - \mathbf{s}_i\cdot\mathbf{s}_j)X_i^{xy,xy}X_j^{xy,xy}. \end{equation}
%and $\tilde{\mathcal{H}}^{\rm p}_{\rm eff} = 0$.
which is a standard result for the Hubbard model with non-degenerate band.

Below we consider the limit of $J_{\rm ex}\ll\Delta_{\rm CF} $ and project the Hamiltonian onto $xz,yz$ doublet (denoting this by a bar), assuming that $t^{xz}_{ij} = t^{yz}_{ij} = t$ for nearest neighbor hopping and zero otherwise.

We get these Hamiltonians in terms of pseudospin operators $\tau^{o}_i = (1/2)\sum_{\sigma,nn'\ne xy}\sigma^{o}_{nn'}c^\dag_{in\sigma}c^{}_{in'\sigma}$~\cite{kugel1972superexchange}, $o = 0,x,y,z$,
\begin{multline}
\bar{\mathcal{H}}^{\rm d}_{\rm eff} = 
 -\frac{2t^2}{U - J_{\rm d}}\left(1 - \frac{J_{\rm d}}{U + \beta J_{\rm d}} \right)\sum_{<ij>}(1/4-\mathbf{s}_i\cdot\mathbf{s}_j) 
 \\
\times\left(\frac14 + \tau^z_i\tau^z_j + \frac{s_{ij}}2(\tau^z_i + \tau^z_j) \vphantom{\frac11}\right),
\end{multline}
\begin{multline}
\bar{\mathcal{H}}^{\mathrm{p}}_{\rm eff} = -\frac{t^2}{U' - J}\sum_{<ij>}
\left(\vphantom{\int}1 - \frac{J/2}{U' + J}
+
2a_0\mathbf{s}_i\cdot\mathbf{s}_j\right)
\\
\times \left(\frac14 - \tau^z_i\tau^z_j \vphantom{\frac11}\right),
\end{multline}
where $s_{ij}=a^{xz}_{ij}-a^{yz}_{ij}$ and we have taken into account that $\tau^0_i = 1/2$~(one electron per site filling). 

Writing down the total effective projected Hamiltonian as 
\begin{equation}\label{eq:H_eff}
\bar{\mathcal{H}}_{\rm eff} = \mathcal{{H}}_{\rm exch}  + \mathcal{{H}}_{\rm so},    
\end{equation}
we neglect the terms of order $\lambda t^2/U^2$. 
%and $\Delta_{\rm CF}t^2/U^2$???. 
This model is well suited to describe  layered perovskites like Sr$_2$VO$_4$ with degenerate orbital states  and large Coulomb interaction. For this compound,  estimations  yield  $U \simeq 20t$,  and $\Delta_{\rm CF}$ varies rather strongly \cite{2024:Igoshev_Chizhov,chizhov2025orbital}.
%$W\ll U$ ~($W$ --- electron band width).
We treat the situation of a  crystal field splitting  $\Delta_{\rm CF} \gg \lambda, t^2/(U - 3J)$. % можно считать выключенной из рассмотрения. 
%Будем сравнивать результаты, получающиеся 

When $\Delta_{\rm{CF}} \sim t$, the occupation of the $xy$ state is small but finite.
For $\Delta_{\rm{CF}}\gg t$, the $xy$ orbital is always unoccupied, so that 
%Let the $xz$ orbital ($yz$) correspond to pseudospin up (down), then
the sums over $m$, $m'$ are in fact taken only over these two orbitals. 
For a half-filled band considered $\tau^0_i = 1/2$.

We apply the Kanamori parametrization $U' = U - 2J$, $J = J_d$~\cite{2013:Georges_Coulomb}. 
\begin{multline}\label{eq:H_SQ_exch}
\bar{\mathcal{H}}_{\rm eff} = NE_0^\beta 
+ \frac12\frac{t^2}{U - J}\sum_{<ij>}
\\
\left(u_z^\beta(J/U)
\tau^z_i\tau^z_j
+v_0^\beta(J/U)\mathbf{s}_i\cdot\mathbf{s}_j
+ 4v_z^\beta(J/U)
\tau^z_i\tau^z_j\mathbf{s}_i\cdot\mathbf{s}_j
\right)
\\
+\frac{t^2}{U - J}\left(1 - \frac{J}{U + \beta J} \right)\sum_{<ij>}s_{ij}\mathbf{s}_i\cdot\mathbf{s}_j 
(\tau^z_i + \tau^z_j).
\end{multline}
where $N$ is number of lattice sites, %$J_{\rm ex} = t^2/(U - J)$ 
$$
E_0^\beta =  -\frac18\frac{zt^2}{U - J}\left(3 + 
J\left(\frac{3}{U - 3J} - \frac{1}{U + \beta J}\right)\right)
$$
is constant contribution to energy,
\begin{multline}
 u_z^\beta(x) = \frac{1}{1-3x} + \frac{x}{1+\beta x}, \,  v_0^\beta(x) = \frac{1-4x}{1-3x} - \frac{x}{1+\beta x},\\
v_z^\beta(x) = \frac{1-2x}{1-3x} - \frac{x}{1+\beta x}.
\end{multline}

It is convenient to introduce %\petr{[???ispravil]}
isospin operators $\hat{I}^{x}_i = 2\tau^z_is^{x}_i$, $\hat{I}^{y}_i = 2\tau^z_is^{y}_i$, $\hat{I}^{z}_i = s^{z}_i$~\cite{Jackeli2009a}. 

Also, in this case it is convenient to employ the quantum number of a \textit{pseudoorbital}: $\zeta=+$ ($-$), if the $z$-projection of the orbital momentum is co-directed (oppositely directed) with the spin $z$-projection. Then we have the basis $\left|\sigma\right>_\zeta \equiv \left|l^{z}=\zeta\sigma\right>\otimes\left|\sigma\right>$, where $|l^{z}=\pm1\rangle = (\mp|yz\rangle -\I |xz\rangle)/\sqrt2$. 
The convenience of this basis is  owing to the simplicity of the $z$-contribution to the spin-orbit Hamiltonian dominating in the case $\Delta_{\rm CF}\gg t^2/(U - 3J)$: the doublet with a fixed $\zeta$ is an eigenspace of $\mathcal{H_{\rm so}}$ with the eigenvalue $-\zeta\lambda/2$. Then we get in this limit
%For the spin-orbit interaction Hamiltonian  with $\Delta_{\rm CF}\gg t^2/U$ we obtain
\begin{equation}\label{eq:H_so_pseudospin}
\mathcal{{H}}_{\rm so} = -\lambda\sum_{i}\hat{I}^{z}_i l^{z}_i = -\frac\lambda2\sum_{i\zeta}\zeta\mathcal{P}^\zeta_i,
\end{equation}
where $\mathcal{P}^\zeta_i$ is the $\zeta$-subspace projection operator at a site~$i$.

Then the Hamiltonian \eqref{eq:H_SQ_exch} can be represented  as:
\begin{multline}\label{eq:H_SQ_exch.final}
\bar{\mathcal{H}}_{\rm eff} =  
\frac12\frac{t^2}{U - J}\sum_{<ij>}\left(u_z^\beta(J/U)
\tau^z_i\tau^z_j +v_0^\beta(J/U)\mathbf{s}_i\cdot\mathbf{s}_j
\right.
\\
\left.
+ v_z^\beta(J/U)
(\hat{\mathbf{I}}^\perp_i\cdot\hat{\mathbf{I}}^\perp_j + 4\tau^z_i\tau^z_j\hat{I}^z_i\hat{I}^z_j)
\right)
+\frac{t^2}{U - J}\left(1 - \frac{J}{U + \beta J} \right)
\\
\times\sum_{<ij>}\frac{s_{ij}}2(\mathbf{s}^\perp_i\cdot\hat{\mathbf{I}}^\perp_j + \hat{\mathbf{I}}^\perp_i\cdot\mathbf{s}^\perp_j + 
2\hat{I}^z_i\hat{I}^z_j(\tau^z_i + \tau^z_j)).
\end{multline}
where the vector notations $\hat{\mathbf{I}}^\perp=(\hat{I}^x,\hat{I}^y,0)$, $\mathbf{s}^\perp=(s^x,s^y,0)$ are used.

It can be easily shown that in the zeroth order in $J/U$, $u^\beta_z(J/U)$ and $v^\beta_{0,z}(J/U)$ become unity and all isospin components correspond to the same exchange integral, i.e., the isospin exchange is completely isotropic.
Also, taking into account $J/U$ terms, the  exchange interaction becomes anisotropic: for the $x,y$ components, it is enhanced by a contribution proportional to $1/4 - \tau^{z}_{i}
\tau^{z}_{j}\ge 0$ (easy-plane anisotropy in the real space).    
%Обменный изоспиновый гамильтониан обладает симметрией???.  

\section{Phases with magnetic and orbital ordering}

\begin{figure}[h!]
\includegraphics[width=.483\textwidth]{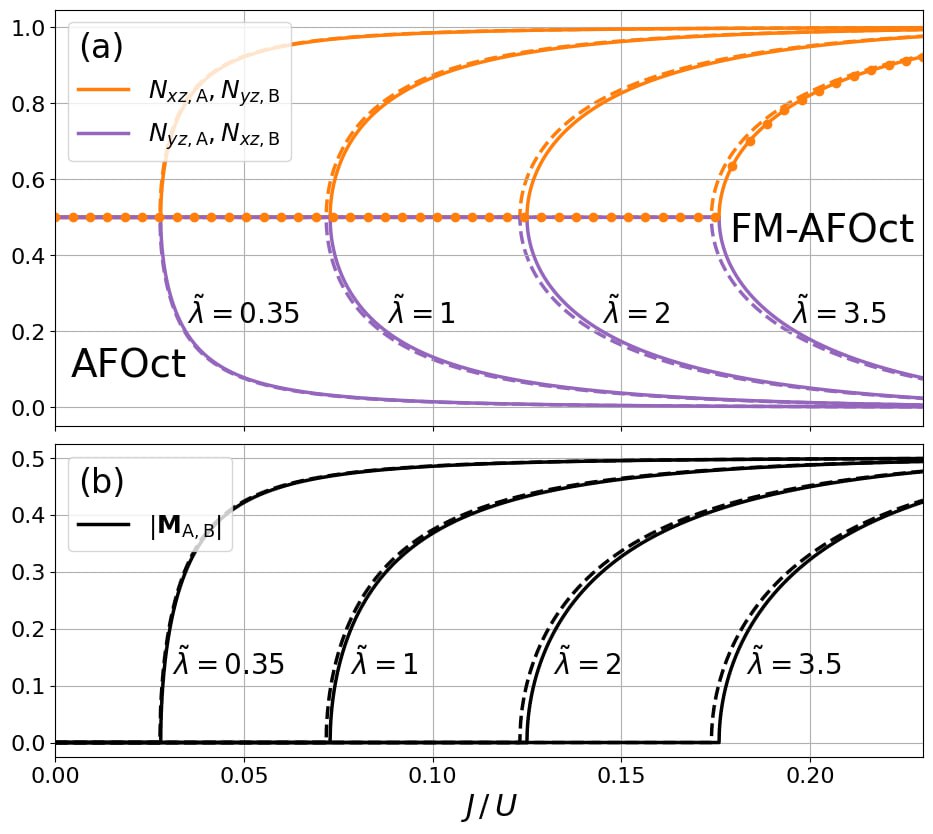}
\caption{
$J$-dependences for the model \eqref{eq:H_eff}: (a) of orbital filling $N_{im} = \langle X_{i}^{mm}\rangle$ of sublattices, $N_{\mathrm{A}xz} = N_{\mathrm{B}yz}$, $N_{\mathrm{A}yz} = N_{\mathrm{B}xz}$; (b)  of absolute value of site magnetization $|\bf{M}_{s}|$. The parameter values $\tilde{\lambda} = \lambda U/t^2 = 0.35, 1.0, 2.0, 3.5$ are presented. Solid lines correspond to $t \ll \Delta_{\rm CF}\ll U$ ($\beta =2$), and dashed lines to  $U \ll \Delta_{\rm CF}$ ($\beta =1$). 
}
\label{fig:three_subplot}
\end{figure}    

To explore the  Hamiltonian \eqref{eq:H_eff} 
%in the following double-sublattice approximation 
we use the trial wave function
\begin{equation}\label{eq:wave_function_ansatz}
    \left|\Psi\right> = \prod_{s}\prod_{i\in s}\sum_{m\sigma}\alpha^s_{m\sigma}c^\dag_{im\sigma}\left|\text{VAC}\right>,
\end{equation}
where $\left|\text{VAC}\right>$ is the empty lattice state (vacuum), $\alpha^s_{m\sigma}$ are the normalization factors $\left(\sum_{m\sigma}|\alpha_{m\sigma}^s|^2 = 1\right)$, $s\in\{{\rm A}, {\rm B}\}$ is a sublattice index.
In this approximation, the problem of finding the ground state of the model is equivalent to the problem of minimizing the  energy $E = \langle\Psi|\mathcal{H}_{\rm eff}|\Psi\rangle$ with respect to the factors $\alpha^s_{m\sigma}$. 
%In this case, $E$ is represented as a real biquadratic form of these coefficients. To optimize this form, the iterative algorithm described in the papers~\cite{2009:algebra,2024:algebra} was used. 
We optimized the corresponding  biquadratic form by employing the iteration algorithm developed in Refs.~\cite{2009:algebra,2024:algebra}.

%Since isospin is analogous to spin, 
The on-site wave function is represented as a linear combination of isospinors:
\begin{equation}\label{eq:psi_start}
\psi = \sum_{\zeta}z_{\zeta}\left|\mathbf{n}_{\zeta}\right>_\zeta,
\end{equation}
where, by analogy with ordinary spinors, a standard parametrization is introduced in the spherical coordinate system
\begin{equation}\label{eq:spinor_repr}
\left|\mathbf{n}_{\zeta}\right>_\zeta  = \cos\left(\frac{\Theta_\zeta}{2}\right) \left|\uparrow\right>_\zeta + \sin\left(\frac{\Theta_\zeta}{2}\right)e^{\I\Phi_\zeta} \left|\downarrow\right>_\zeta,
\end{equation}
%суммирование ведется по индексу псевдоорбитали, а $z_\sigma$ --- коэффициенты, 
$\mathbf{n}_\zeta$ 
being unit vector indicating the isospin direction, $z_{+}=\cos\left(\frac{\theta}{2}\right)$, $z_{-}=\sin\left(\frac{\theta}{2}\right)e^{\I\phi}$.
%, $0 \le \Theta_\zeta,\theta\le\pi$, $-\pi<\Phi_\zeta,\phi\le\pi$. 
Thus, the parameter $\theta$ determines the filling of the pseudo-orbitals ``$\pm$''.

% В общем случае для произвольной одноэлектронной функции $\psi$, см. уравнение \eqref{eq:psi_start}

In the general case of an arbitrary one-electron function $\psi$, we have for the averages
\begin{eqnarray}
\langle\psi|\mathbf{s}|\psi\rangle &=& \frac{\mathbf{e}_z}2\sum_\zeta |z_\zeta|^2(\mathbf{n}_\zeta \mathbf{e}_z) + \frac12\sum_\zeta z^*_{\overline\zeta}z_{\zeta}\mathbf{n}^\perp_\zeta\langle\mathbf{n}_{\overline\zeta}|\mathbf{n}_\zeta\rangle, \nonumber \\
\langle\psi|\mathbf{\hat{I}}|\psi\rangle &=& \frac12\sum_\zeta |z_\zeta|^2\mathbf{n}_\zeta, \, 
\langle\psi|\tau^{z}|\psi\rangle = \frac12 \sum_\zeta z^*_{\overline\zeta}z_{\zeta}\langle\mathbf{n}_{\overline\zeta}|\mathbf{n}_\zeta\rangle.\nonumber
\end{eqnarray}

The ground state for a sublattice can be characterized by one angle-dependent unit vector $\mathbf{n} = \mathbf{n}_{\pm}$, coinciding with the isospin direction. We denote such a state as $\psi_\mathbf{n}(\theta,\phi)$.

The following states are possible:

(i) FM-AFO$_{xz/yz}$ (ferromagnetic---antiferrorbital phase) --- a phase with saturated ferromagnetic spin ordering and antiferromagnetic orbital ordering concerning the $xz$ and $yz$ orbitals~\cite{kugel1982jahn}. 
%In Ref.~\cite{2024:Igoshev_Chizhov}, such a phase was considered for the case where the average spin is directed along the $Oz$ axis (we will call it FM-AFO$^\parallel_{xz/yz}$); 
It is realized only for $\lambda=0, \, J > 0$. 
%and will not be considered below.
%As $\lambda$ increases, the wave function becomes admixed with an orbital of the opposite sublattice, while the orbital and spin degrees of freedom remain entangled.

%В таком непрерывном продолжении, рассмотренном в работе \cite{2024:Igoshev_Chizhov}, средний спин направлен вдоль оси $Oz$, называть такую фазу будем FM-AFO$^\parallel_{xz/yz}$. 

(ii) AFOct (antiferrooctupole phase) --- a phase with hidden antiferromagnetic orbital  ordering \cite{Jackeli2009a}. 
%in the limit of a large crystal field, which corresponds to our model at $J = 0$. For any value of the parameters in this phase, for each sublattice $\theta=0$ ($\phi$, $\Theta_-$, $\Phi_-$ are not determined), $\Theta_+=\frac\pi2$. This means that the electron occupies the pseudoorbital ``$+$'' and the isospin lies in the $xy$ plane. $e^{\I\Phi_+}$ has different signs on the sublattices; there is degeneracy in the angle $\Phi_+=\Phi$, which determines the direction of the spins and isospins in the $xy$ plane. Thus, the wave functions of the sublattices in this phase can be written as $\left|\psi^{\rm AFOct}_{\rm A,B}\right\rangle = \left({\left|\uparrow\right\rangle_+\pm e^{\I\Phi}\left|\downarrow\right\rangle_+}\right)/{\sqrt2}$  with opposite signs on the sublattices. 
In this phase, the dipole spin and orbital moments are absent on each sublattice, and  only  four components of the octupole moment, $T^\alpha_{x,y}=\pm 15n_{x,y}/8$ and $T^\beta_{x,y}=3\sqrt{15}n_{x,y}/8$~\cite{2009:Santini}, remain nonzero, $\pm$ signs corresponding to $x,y$ components. They have opposite signs on the sublattices, meaning that antiferrooctupole order is realized.
% $\frac12\left(-\left(\left|yz,\uparrow\right>+\I\left|xz,\uparrow\right>\right)\pm\left(\left|yz,\downarrow\right>-\I\left|xz,\downarrow\right>\right)\right)$

(iii) FM-AFOct is a ferromagnetic phase (ferrioctupole with respect to the octupole order), with orbital occupations being intermediate between AFOct and the saturated   FM-AFO$_{xz/yz}$ phase, i.e., there is a weak antiferroorbital order. 
%The wave functions also depend on $J$ and $\lambda$, but, unlike FM-AFO$^\parallel_{xz/yz}$, the orbital and spin degrees of freedom in the FM-AFOct phase are entangled. For any value of the parameters in the FM-AFOct phase, 
Here $\Theta_\pm = \pi/2$ for both sublattices, $\theta$~coincides for sublattices, $e^{\I\phi}$ and $e^{\I\Phi_+}=e^{\I\Phi_-}=e^{\I\Phi}$ have different signs on the sublattices, $e^{\I\phi}=\pm1$, similarly to AFOct, there is a degeneracy in the angle $\Phi$; $\theta$ depends on $J$, $\lambda$. Thus, the wave functions read
%\begin{equation}
$\psi^{\rm FM-AFOct}_{\rm A,B} = \frac1{\sqrt2}\left(\cos\frac{\theta}{2}\left(\left|\uparrow\right>_+ \pm e^{\I\Phi}\left|\downarrow\right>_+\right) + \sin\frac{\theta}{2}\left(e^{\I\Phi}\left|\downarrow\right>_-\pm\left|\uparrow\right>_-\right)\right)$
%\end{equation}
with different signs on the sublattices A, B. In the limit of a small  $\lambda\ll t^2J/U^2$, ferromagnetism in the $xy$ plane becomes saturated.
%as a special case of the FM-AFOct phase at $\theta \sim \pi/2$, a saturated ferromagnetic phase with an average spin lying in the $xy$ plane is realized, which we will call FM-AFO$^\perp_{xz/yz}$.
%Для тех состояний, которые реализуются, мы получили следующее:

It is possible to obtain a simple analytical expression for the phase boundary. 
Assuming that $M^z_s = 0$ where $\mathbf{M}_s=\langle \mathbf{s}_s\rangle$~($\hat{\mathbf{I}}_s=\langle \mathbf{\hat{I}}_s\rangle$) is a vector of magnetic moment (isospin) on sublattice $s$, for $E_{\rm exch}=\langle\bar{\mathcal{H}}_{\rm eff} \rangle/N$ in the two-sublattice approximation  from the expression \eqref{eq:H_SQ_exch.final} we obtain for the exchange energy
\begin{multline}
E_{\rm exch}  = \frac{2t^2}{U-J}
\left[u_z^\beta\left(\frac JU\right) T_{\rm A}T_{\rm B}
\right.
\\
\left.
+ v_0^\beta\left(\frac JU\right)\mathbf{M}_{\rm A}\mathbf{M}_{\rm B}+ v_z^\beta\left(\frac JU\right)\hat{\mathbf{I}}_{\rm A}\hat{\mathbf{I}}_{\rm B}
\right]
\end{multline}
%\end{equation}
where $T_s=\langle \tau^{z}_s\rangle=\frac12\left(N_{s,xz}-N_{s,yz}\right)$ is average  pseudospin, $N_{s,m}=\sum_\sigma\langle c^\dag_{sm\sigma}c^{}_{sm\sigma}\rangle$ is  occupation of orbital $m$ at any site of  sublattice $s$.
Similarly, we define the average energy for the operator $\mathcal{{H}}_{\rm so}$: 
%per site: $E_{\rm so} = \langle\mathcal{{H}}_{\rm so}\rangle/N$. In the two-sublattice approximation
\begin{equation}\label{eq:E_so}
E_{\rm so} = -\frac\lambda4\sum_{s\zeta}\zeta\langle\mathcal{P}^\zeta_s\rangle.
\end{equation}
%$\bar{{\rm A}} = {\rm B}, \bar{{\rm B}} = {\rm A}$.
Three terms proportional to $u_z^\beta$, $v_0^\beta$, and $v_z^\beta$ describe exchange interaction: for pseudospins,  spins, and isospins. The coupling constants depend differently on $J$: 
%see Fig.~\ref{fig:contrib}(b): 
as $J$ increases, pseudospin exchange increases faster than isospin exchange, while spin exchange falls changing sign at $J/U \simeq 0.24$.

The FM-AFOct phase is characterized by a significant total spin angular moment in the $xy$ plane, which increases with $J$ above the critical value, being equal for both sublattices, see Fig.~\ref{fig:three_subplot}(b). 
One can see that the increase is especially  rapid for small $\lambda$.
Also, the picture depends very weakly on the crystal field splitting, which is important when treating experimental data.

Furthermore, the AFOct and FM-AFOct phases can be described by the isospin vector, which has a maximum average value of 1/2, unlike the average spin. Its magnitude is conserved for both phases, and on one sublattice it is codirectional with the spin angular momentum, while on the other it is oppositely directed.
%, see Figure~\ref{fig:three_subplot}(c).

\begin{figure}[h!]
\includegraphics[width=.483\textwidth]{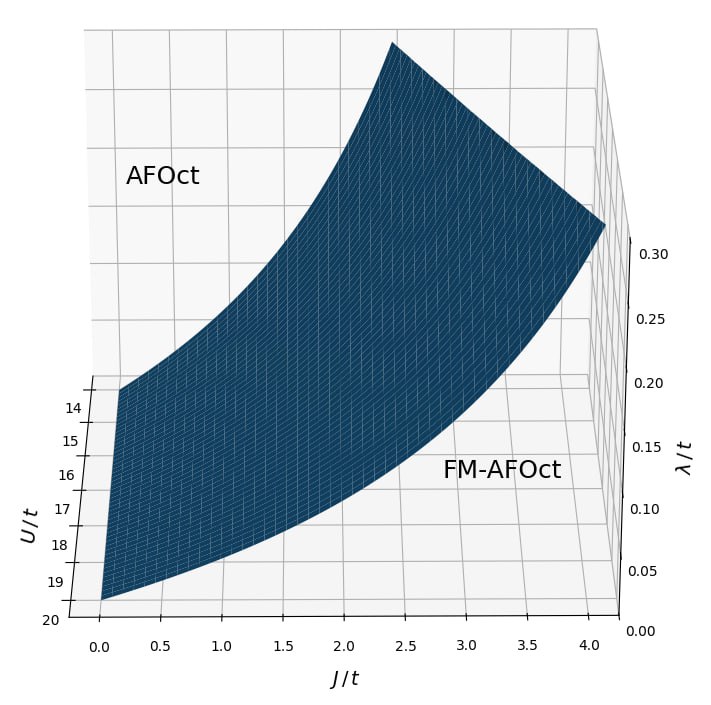}
\caption{
The phase diagram in $J-U-\lambda$ variables for the model \eqref{eq:H_eff}  with 
%$\Delta_{\rm CF}=t$ and 
$t \ll \Delta_{\rm CF}\ll U$.
AFOct is a phase with hidden antiferrooctupole ordering~\cite{Jackeli2009a}, FM-AFOct is a ferromagnetic phase.
}
\label{fig:phase_diagram}
\end{figure}

\begin{figure}[h!]
\includegraphics[width=.483\textwidth]{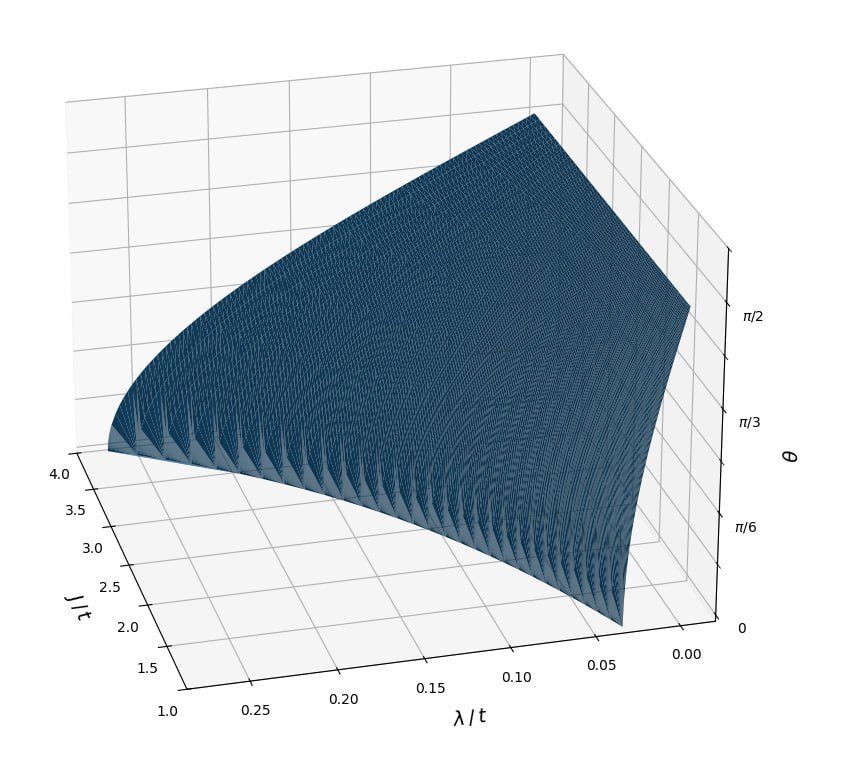}
\caption{
The dependence of the pseudoorbital-space polar angle $\theta$ in $J-\lambda$ variables  for the FM-AFOct ferromagnetic phase  with $t \ll \Delta_{\rm CF}\ll U$, $U = 20t$.
}
\label{fig:phase_diagram}
\end{figure}

\section{Analytical solution}

In the state $\psi = \psi_\mathbf{n}(\theta,\phi)$ we obtain $\hat{\mathbf{I}}(\psi_\mathbf{n}) = \frac{\mathbf{n}}2$, $\mathbf{M}(\psi_\mathbf{n}) = \frac{\mathbf{n}}2\sum_\zeta {z}^*_{\overline\zeta}z_\zeta=\pm\frac{\mathbf{n}}2\sin\theta$, $T(\psi_\mathbf{n}) = \frac12 \sum_\zeta {z}^*_{\overline\zeta}z_\zeta=\pm\frac12\sin\theta$, $\sum_\zeta\zeta\langle\psi|\mathcal{P}^\zeta|\psi\rangle = \sum_\zeta |z_\zeta|^2=\cos\theta$.
The site magnetic moment occurs simultaneously with the filling of the pseudo-orbital ``$-$'', i.e., with the appearance of orbital ordering --- the difference in orbital filling coincides with the magnitude of the magnetic moment.
After substitution into the total energy $E = E_{\rm exch} + E_{\rm so}$ we get
%$E=E_{\rm exch,1}+E_{\rm exch,2}+E_{\rm exch,3}+E_{\rm so}$: 
%\begin{equation}
\begin{multline}
E
%=-\frac\lambda4\sum_{s}\sigma_s^{(z)}+2J_{\rm ex}\left(u_z\left(\frac JU\right)\tau^{(z)}_{\rm A}\tau^{(z)}_{\rm B}+v_0\left(\frac JU\right)\mathbf{M}_{\rm A}\mathbf{M}_{\rm B}+v_z\left(\frac JU\right)\hat{\mathbf{I}}_{\rm A}\hat{\mathbf{I}}_{\rm B}\right)+E_0
=-\frac\lambda2\cos\theta
-\frac12 \frac {t^2}{U-J}[(u_z^\beta(J/U)
-v_0^\beta(J/U))\sin^2\theta \\
+ v_z^\beta(J/U)].
\end{multline}
% \\=-\frac\lambda2\cos\theta-J_{\rm ex}\left(\frac JU\left(\frac2{1-3J/U}+\frac1{1+2J/U}\right)\sin^2\theta+\frac12v_z\left(\frac JU\right)\right)+E_0
% \\=-\frac\lambda2\cos\theta-J_{\rm ex}\left(\frac{J/U(3+J/U)}{(1-3J/U)(1+2J/U)}\sin^2\theta+\frac12v_z\left(\frac JU\right)\right)+E_0.
%\end{equation}
Minimizing the  energy results in
%\begin{equation}\label{eq:cos_theta} \cos\theta_{\rm c}(\lambda, J)=\frac{\lambda}{4J_{\rm ex}}\cdot\frac1{J/U}\cdot\frac{(1-3J/U) 1+2J/U)}{3 + J/U}. \end{equation}
%\begin{equation} !!\lambda_{\rm c}(J) =4J_{\rm ex}\frac{J/U(3+J/U)}{(1-3J/U)(1+2J/U)}.\end{equation}
%\begin{equation}\label{eq:cos_theta} \cos\theta(\tilde\lambda, \frac JU)=\frac{\tilde\lambda}{4}f(\frac JU), \, f(x)=\frac{(1-3x)(1+2x)}{x(3 + x)} \end{equation}
\begin{eqnarray}
    \label{eq:cos_theta}
\cos\theta(\tilde\lambda, \frac JU)=\frac{\tilde\lambda}{4}f_\beta(\frac JU), \\ f_\beta(x)=\frac{(1-x)(1-3x)(1+\beta x)}{x(3 + (2\beta - 3)x)} 
\end{eqnarray}
with $\tilde{\lambda}= \lambda U/t^2$ the dimensionless spin-orbit coupling parameter.
Note that for  the exotic situation of negative $\lambda$ we can replace in (\ref{eq:cos_theta}) $\lambda \rightarrow -\lambda$, $\cos \theta \rightarrow \cos (\pi- \theta)$, so that $\theta>\pi/2$.

For any $U$, $J$, $\lambda$ with $0<x=J/U<1/3$, the function $f_\beta(x)$  varies monotonically from $+\infty$ to zero. 
Thus, there is a quantum phase transition: if $J/U$ exceeds the critical value, $\cos\theta$ behaves according to the formula \eqref{eq:cos_theta}, if less --- $\cos\theta=1$. The phase boundary is described by 
%(рисунок~\ref{fig:phase_diagram}):
% \begin{equation}
% \frac{\lambda/t\cdot U/t}{4}=\frac{J/U(3+J/U)}{(1-J/U)(1-3J/U)(1+2J/U)}.
% \end{equation}
\begin{equation}
\lambda_{\rm c}(J)% =4J_{\rm ex}\frac{J(3U+J)}{(U-3J)(U+2J)}
= \frac {4 t^2J}{U-J}\frac{3U+(2\beta-3) J}{(U-3J)(U+\beta J)}.
\end{equation}
The corresponding magnetic phase diagram for $\beta=2$ is presented in Fig.2.

Performing the expansion $\cos\theta = 1 - \theta^2/2$ we obtain the dependence of the order parameter $\theta$ in the vicinity of the critical point
\begin{equation}
\theta \simeq \sqrt{2(1-\lambda/{\lambda}_{\rm c}(J))}.
\end{equation}

For given $U$ and $\lambda$, the phase transition point is determined by the  cubic equation $\tilde{\lambda}f_\beta(x)=4$ with $x=J/U$,   
\begin{eqnarray} 
6\tilde{\lambda}x^3 - \left((4\beta-3)\tilde{\lambda}+8\beta -12\right) x^2 \nonumber
\\- \left((4-\beta)\tilde{\lambda}+12\right) x +\tilde{\lambda} = 0.
\end{eqnarray}
%\begin{equation}3 \beta\tilde{\lambda}x^3 - \left(5\tilde{\lambda}+4\right) x^2 - 2\left(\tilde{\lambda}+6\right) x +\tilde{\lambda} = 0.\end{equation}
%[Since $f'(x) = 18\tilde{\lambda}x^2 - 2\left(5\tilde{\lambda}+4\right)x - 2(\tilde{\lambda}+6) < 18\tilde{\lambda}x^2 - 2(\tilde{\lambda}+6) < 2\tilde{\lambda}- 2(\tilde{\lambda}+6) < 0$ there is only one solution of Eq.~(16)  in the interval $0<x<1/3$. }]
This equation has one  root $x_c$ in the interval $0<x<1/3$. 
In particular, for $\beta=2$ (charge-transfer regime) we have 
%which is given by
\begin{eqnarray}
x_c^{-1}&=&-2{\sqrt {Q}}\cos \left(\alpha +{ {2}}\pi/3 \right)+{2(1+6/\tilde{\lambda})/{3}}, \nonumber\\
\alpha&=&\arccos(R/Q^{3/2})/3,\nonumber\\
R&=&28/27 - 50 \tilde{\lambda}^{-1}/3 -40 \tilde{\lambda}^{-2} - 64 \tilde{\lambda}^{-3},\nonumber\\
Q&=&19/9 + 20 \tilde{\lambda}^{-1}/3 + 16 \tilde{\lambda}^{-2}.
\end{eqnarray}

Near the critical value we have
\begin{equation}
\theta \simeq \sqrt{\tilde{\lambda}f'_\beta(x_c)(J/U-x_c)/2}.
\end{equation}
For small $J$ we obtain $J_{\rm c}={\tilde{\lambda}}U/12$,
\begin{equation}
\theta \simeq \sqrt{2(J/J_{\rm c}-1)}.
\end{equation}

Fig.~\ref{fig:phase_diagram} shows the behavior of the angle $\theta$ in the variables $J-\lambda$ for a fixed $U$.

\section{Conclusions}

We have treated the ordered states and phase transitions in an effective anisotropic Heisenberg model constructed for the $t_{\rm 2g}$ states with an exact account of the crystal field, the ground state phase diagram being studied analytically.
A solution is described in terms of an isospinor wave function for eigenfunctions of the spin-orbit coupling operators corresponding to two pseudo-orbitals.
The nature of long-range order is governed by the competition of Hund's exchange and spin-orbit coupling, the dependence on the crystal field splitting being rather weak.

The phase diagram in the variables $J-\lambda$ shows that for a given value of $\lambda$, a small value of $J$ leads to the formation of a hidden (antiferrooctupole) order with equal occupation of the two orbitals $xz$ and $yz$, and  increasing $J$ leads to a continuous transition into a  ferromagnetic phase combined with orbital ordering, and its amplitude is equal to the value of the partially suppressed ferromagnetic moment. 
This state is intermediate between well-known ferromagnetic antiferroorbital and hidden orders~\cite{kugel1982jahn}.
The spin-orbit interaction destroys the saturated ferromagnetic state caused by the Hund exchange, forming a two-sublattice state with hidden orbital and spin order.

The results obtained provide a qualitative description of  orbital ordering in the layered perovskite Sr$_2$VO$_4$. 
We expect that this compound, exhibiting ferromagnetic ordering with reduced moment, is close to the phase boundary separating AFOct and FM-AFOct phases. 
Our approach can be applied to investigate magnetic phase transitions in other compounds with various two- and three-dimensional lattices.

%Acknowledgments.
The authors thank S.~V.~Streltsov and A.~N.~Ignatenko for helpful discussions.
This work was completed within the  state assignment from the Ministry of Science and Higher Education of the Russian Federation for the Institute of Metal Physics. 

\bibliography{igoshev_pa.bib}
\end{document}